\documentclass{iau} 

\usepackage{graphicx}

\title[IAUS342: Black hole mass measurements in AGN] %% give here short title %%
{Black hole mass measurements in AGN: \\ Polarization in broad emission lines}

\author[Popovi\'c et al.]   %% give here short author list %%
{Luka \v C. Popovi\'c$^1$, Victor L. Afanasiev$^2$, \and Djordje Savi\'c$^1$
 }

\affiliation{$^1$
Astronomical Observatory,  11160 Belgrade, Serbia,
\\ email: {\tt lpopovic@aob.rs; djsavic@aob.rs} \\[\affilskip]
$^2$Special Astrophysical Observatory of the Russian AS,\\ Nizhnij Arkhyz, 
Karachaevo-Cherkesia 369167, Russia
\\email: {\tt vafan@sao.ru}}

\pubyear{2018}
\volume{xxx}  %% insert here IAU Symposium No.
\jname{IAUS342: Perseus in Sicily}
\editors{Editros}

\begin{document}

\maketitle

\begin{abstract}
We present a new method for supermassive black hole (SMBH) mass measurements in Type 1 active 
galactic nuclei (AGN) using polarization angle across broad lines. 
This method gives measured masses which are in a good agreement with reverberation 
estimates. Additionally, we explore the possibilities and limits of 
this method using the STOKES radiative transfer code taking 
a dominant Keplerian motion in the broad line region (BLR). 
We found that this method can be used for 
the direct SMBH mass estimation in the cases when in addition to the Kepler motion,
radial inflows or 
vertical outflows are present in the BLR. Some advantages of the method are discussed.
%% add here a maximum of 10 keywords, to be taken form the file <Keywords.txt>
\end{abstract}

\firstsection % if your document starts with a section,
              % remove some space above using this command.
\section{Introduction}

Supermassive Black Holes (SMBHs, $10^6-10^9\ M_\odot$) are supposed to reside  in 
the bulges of spiral and elliptical galaxies, and they are in action to shape 
their cosmic environment, i.e. it seems that there is a connection between the central SMBH and 
the host galaxy structure. The close 
connection between the formation and evolution of galaxies and of their central SMBHs involves a 
variety of physical phenomena of great relevance in modern astrophysics (see e.g. Heckman
\& Best 2014).
The parameters which define a SMBH are mass, spin and electricity, where 
mass of SMBH is the most important and probably is in the correlation with the galaxy bulge mass.
Therefore,  one of the most important
issue in astrophysics today is to measure masses of SMBHs.

One of the most powerful objects in the Universe are active galactic nuclei (AGNs) which emit
huge amount of  energy that is created around SMBHs. The emission gas is very close to the 
central SMBH, and therefore the gas kinematics is directly influenced by the mass of the 
central black hole. Especially if the gas emits  the broad lines from the so called 
broad line region (BLR).  The width of 
lines emitted from the BLR corresponds to the rotational velocity
that can be used for estimation of the 
black hole masses in AGNs.

In principle there are several methods for the SMBH mass measurements 
(see Peterson 2014), out of which the reverberation
method is the one often used in the case of AGNs. Using this method one can estimate 
the SMBH mass as:

$$M_{\rm BH}=f{R_{\rm BLR} \sigma_V^2\over G},$$
where $R_{\rm BLR}$ is the photometric dimension of the BLR  obtained from 
the reverberation mapping, and $\sigma_V$ is the corresponding  orbital velocity that can be
estimated from the broad line widths. $G$ is the gravitational constant and
the virial factor $f$ depends on the BLR inclination
and geometry which are unknown and, consequently there is a problem to accurately  determine the 
virial factor (Peterson 2014).

On the other side, the polarization in the broad lines can be very useful for 
investigation of the nature of the BLR (Smith et al. 2002, 2004, 2005;
Afanasiev et al 2014; Afanasiev \& Popovi\'c 2015, Afanasiev et al. 2018 etc.).
The polarization in the broad lines, especially the polarization angle ($\varphi$) 
 that  shows horizontal S shape across the broad line profile, which indicates a
dominant Keplerian-like motion in the BLR, and dominant equatorial scattering as a 
polarization mechanism in the broad lines (see Afanasiev \& Popovi\'c 2015).

In this contribution we shortly describe the method and give some overview of the 
numerical tests of the validity (limitations) of the method 
which have been performed by STOKES code (see Savi\'c et al. 2018)

\begin{figure}[b]
% \vspace*{-2.0 cm}
\begin{center}
 \includegraphics[width=12cm]{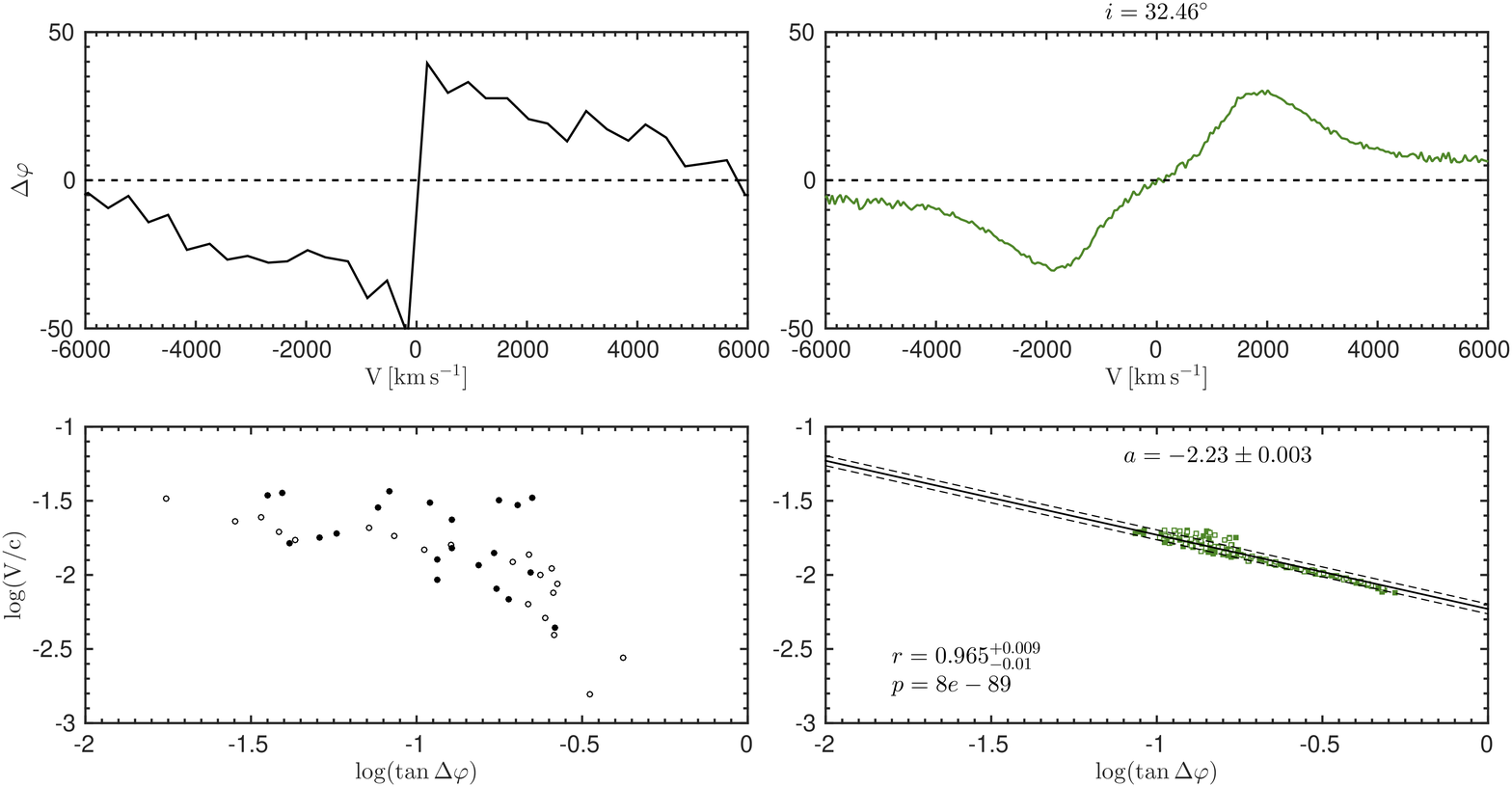} 
 %\includegraphics[width=6cm]{NGC4151.eps}
 % \vspace*{-1.0 cm}
 \caption{The observed (left) and modeled (right) polarization angle changes 
 across the H$\alpha$ line profile (panels up) 
 for NGC 4151 and $\log({V\over c})$ vs. $\tan(\Delta\varphi)$  (panels down) relationship 
(in more detail see Savi\'c et al. 2018).}
   \label{fig1}
\end{center}
\end{figure}

\section{Polarization method for SMBH measurements}

The method is described in more details in \cite{ap15}. We used the fact that in the case
of equatorial scattering, the polarization angle ($\varphi_i$)   is connected with the 
distance ($R_i$) of an emitting cloud (see Fig. 1 in Afanasiev \& Popovi\'c 2015) as 
$$\tan\varphi_i\sim R_i,$$
that in the case of a Keplerian like motion can be connected with the velocity 
($V_i=(\lambda_i-\lambda_0)/\lambda_0$, where $\lambda_0$ is the transition wavelength)
of the emitting
cloud (i.e. $\lambda_i$ is the wavelength emitted by the cloud) as
$$V_i=\sqrt{GM_{\rm BH}\over{R_i}},$$
where  $M_{\rm BH}$ is the mass of SMBH.

It is not difficult to obtain that $R_i= R_{sc}\cdot\tan\varphi_i,$ where $R_{\rm sc}$ is the the 
distance between the SMBH and the scattering region (supposed to be in the inner part of
the torus, see Fig. 1 in Afanasiev \& Popovi\'c 2015). 
Considering the above relations  one can obtain the relationship between the velocity ($V_i$)
 and 
$\tan\varphi$ as:

$$\log({V_i\over c})=a-0.5\cdot \log(\tan(\Delta\varphi_i)),$$
where $c$ is the speed of light, and constant $a$   depends on the SMBH 
mass ($M_{\rm BH}$) as

$$a=0.5\log\bigl({{GM_{\rm BH} \cos^2(\theta)}\over{c^2R_{\rm sc}}}\bigr),$$
that can be used for the SMBH mass measurements. 

We show in \cite{ap15} that 
the method gives SMBH mass estimates which can be compared with the reverberation mapping results.
The method seems to be very perspective, since, 
as it can be seen in Fig. 1 (left panel),  very often we can observe  a horizontal 
S-shaped polarization angle  in the case of Type 1 AGNs (AGNs with the 
broad emission lines, see  Afanasiev et al. 2018).

\section{Modeling and method validity and limitations}

Following the method described by \cite{ap15}, 
we theoretically model the broad line polarization due to equatorial scattering. 
We assume that the unpolarized lines are emitted from the disk-like BLR with dominant
Keplerian motion and scattered by free electrons at the inner part of the dusty torus.
We used 3D Monte Carlo radiative transfer code \textsc{STOKES} 
(see, e.g. Goosmann \& Gaskell 2007, Marin et al. 2015, Marin 2018, Rojas Lobos et~al. 2018) 
 The size of the BLR was obtained from the reverberation mapping with the outer edge set due to 
 dust sublimation. Model parameters of the BLR and the scattering region (SR) are 
 described in more details in \cite{sa18}, and here will not be repeated.
 The models are showing that the mass estimates with the polarization method
 are in the frame of 10\% error-bars in the case that the 
 SR distance which is twice the size of the 
 outer limit of the BLR. The largest number of photons is  scattered only once from the 
 inner edge of the SR. The number of multiple scattering events is negligible and 
 the assumption of a single scattering approximation is valid. In Fig. 1 (right panel) we 
 show the modeled polarization angle across the broad H$\alpha$ profile for NGC 4151, and 
 as one can see from Fig 1, the modeled profile is very similar to the observed one. Also
 the relationship between velocity  and $\tan\varphi$ across the line profile can  be 
 reproduced (Fig 1. bottom panels).
 This indicates that the single element equatorial approximation is valid.
 
 Additionally, we considered the possible inflows/outflows in the BLR 
 (see Savi\'c et al. 2018) and we  have shown that the method for 
 SMBH mass measurements can be used when the contribution of the
 inflows/outflow is much smaller
 in comparison with Keplerian motion. Therefore, this method could potentially be used for highly 
 ionized lines such as C III] and C IV that are observed  in the optical domain  for 
 high-redshifted quasars. 

For high  inclinations, the polar scattering becomes dominant and we have Type-2 objects 
for which the method is no longer valid. For face-on AGNs, the optical polarization is usually 
much lower than 1\% (Smith et al. 2002, Marin 2014) and the amount of interstellar polarization 
can dominate  the equatorial scattering induced polarization in the innermost part of the AGNs. 
The variability of Type-1 AGNs must also be taken into account. The flux variations of the 
continuum and  broad lines can be up to a factor of 10 or greater between minimum and 
maximum activity state (Shapovalova et al. 2008, 2010).
When observed in the minimum activity (up to Type-2), the absence or the weak 
flux in the BLRs could not be detected and the method cannot be used.

\section{Discussion and conclusions}

In this paper we shortly describe the polarization method for the SMBH mass measurement using the 
shape of polarization angle across the broad line profiles (Afanasiev et al.
2014, Afanasiev \& Popovi\'c 2015, Savi\'c et al. 2018). We found that our method gives
very good results in comparison with the reverberation one.

As conclusions we can list the advantages and disadvantages of the method. First,
let us  mention the advantages, which  are:
i) the method does not need assumption of virialization (as in the case of the reverberation one), 
since the horizontal S-like shape of polarization angle across the broad line indicates presence 
of the Keplerian-like motion (see Fig. 1); ii) the method is not time consuming, i.e.
from one epoch  observation of
broad line, one can obtain the SMBH mass; iii) the method can be applied to all broad lines from 
the UV to the optical, i.e. one can measure the SMBH mass of different redshifted AGNs in the same
(consistent) way. 

On the other hand, there are some disadvantages of the method which  are: 
i) the polarization in the broad line of 
AGNs is not so big, therefore the effect can be hidden by the other (additional) effects of 
polarization or depolarization; ii) in the case of very strong outflow and inflows, the effect 
cannot be clearly seen, and could not be used for the SMBH mass
determination and iii) there can be 
a problem with the estimation of  distance of the equatorial scattering region ($R_{\rm sc}$).

Finally, let us conclude that the method is very useful and it represents a new method 
for the SMBH mass determination that can be used also in the case of high redshifted AGNs.

\

\

{\bf Acknowledgement.} 
 This work was supported by the Russian
Foundation for Basic Research ({project N15-02-02101}) and
the Ministry of Education, Science and Technological Development (Republic of Serbia) through
the project Astrophysical Spectroscopy of Extragalactic Objects
(176001).

\end{document}